\documentclass[twocolumn,10pt]{Paper}

\usepackage{epsfig} 
\usepackage{float}
\usepackage{xcolor}

\newcommand{\cc}[1]{\textcolor{red}{#1}}

\title{Viscoelastic Metamaterials}

\author{David M.J. Dykstra
    \affiliation{
	Institute of Physics\\
	University of Amsterdam\\
        Amsterdam, 1098 XH\\
        Netherlands\\
    E-mail: d.m.j.dykstra@uva.nl
    }
}

\author{Joris Busink
    \affiliation{
	Institute of Physics\\
	University of Amsterdam\\
        Amsterdam, 1098 XH\\
        Netherlands\\
    }
}
\author{Bernard Ennis    
	\affiliation{
	Materials Design Department\\
	Tata Steel Europe R\&D\\
        Ijmuiden, 1970 CA\\
        Netherlands\\
		}
}
\author{Corentin Coulais
    \affiliation{
	Institute of Physics\\
	University of Amsterdam\\
        Amsterdam, 1098 XH\\
        Netherlands\\
        E-mail: coulais@uva.nl
    }
}

\begin{document}

\maketitle

\begin{abstract}
{\it Mechanical metamaterials are artificial composites with tunable advanced mechanical properties. Particularly interesting types of mechanical metamaterials are flexible metamaterials, which harness internal rotations and instabilities to exhibit programmable deformations. However, to date such materials have mostly been considered using nearly purely elastic constituents such as neo-Hookean rubbers. Here we explore experimentally the mechanical snap-through response of metamaterials that are made of constituents that exhibit large viscoelastic relaxation effects, encountered in the vast majority of rubbers, in particular in 3D printed rubbers. We show that they exhibit a very strong sensitivity to the loading rate. In particular, the mechanical instability is strongly affected beyond a certain loading rate. We rationalize our findings with a compliant mechanism model augmented with viscoelastic interactions, which captures qualitatively well the reported behavior, suggesting that the sensitivity to loading rate stems from the nonlinear and inhomogeneous deformation rate, provoked by internal rotations. 
Our findings bring a novel understanding of metamaterials in the dynamical regime and opens up avenues for the use of metamaterials for dynamical shape-changing as well as vibration and impact damping applications.}
\end{abstract}

\section{Introduction}
Mechanical metamaterials exhibit a plethora of exotic mechanical responses. 
Static responses of interest span a wide range of tunable behavior, such as auxetic \cite{lakes_foam,bertoldi_negative,babaee_3D}, programmable \cite{florijn1,florijn2}, shape-changing\cite{overvelde_reconf, coulais_multi}, non-reciprocal \cite{coulais_recip} to chiral responses \cite{frenzel_twist}, often by harnessing nonlinear mechanics and snap-through instabilities \cite{florijn1,florijn2,rafsanjani_bistable,chi_multistable,restrepo_phase}. Interesting dynamical responses include shock absorption \cite{frenzel_microlat,shan_trapping,correa_honeycombs,schaedler_metallic} and soliton propagation \cite{deng_soliton,deng_gaps} and transition waves \cite{Raney_propagation,nadkarni_unidir}. Importantly, a compliant mechanism framework \cite{florijn1,rafsanjani_bistable,Coulais_IJSS2016,overvelde_reconf,coulais_recip,Raney_propagation,nadkarni_unidir,coulais_charac,deng_soliton,deng_gaps} is often employed to capture qualitatively the mechanical response and to explore the design space. 

However so far, the effect of the constitutive materials' dissipation has been largely overlooked for nonlinear metamaterials. As a matter of fact, extensive care has often been devoted in the constitutive materials' choice to avoid strong dissipative effects. Viscoelastic effects have been considered in dynamic regimes, but mainly for linear metamaterials \cite{wang_kagome,parnell_soft} and single snap-through elements \cite{gomez_dynamics,arrieta_snap}.

\begin{figure}
\includegraphics[width=1.0\columnwidth]{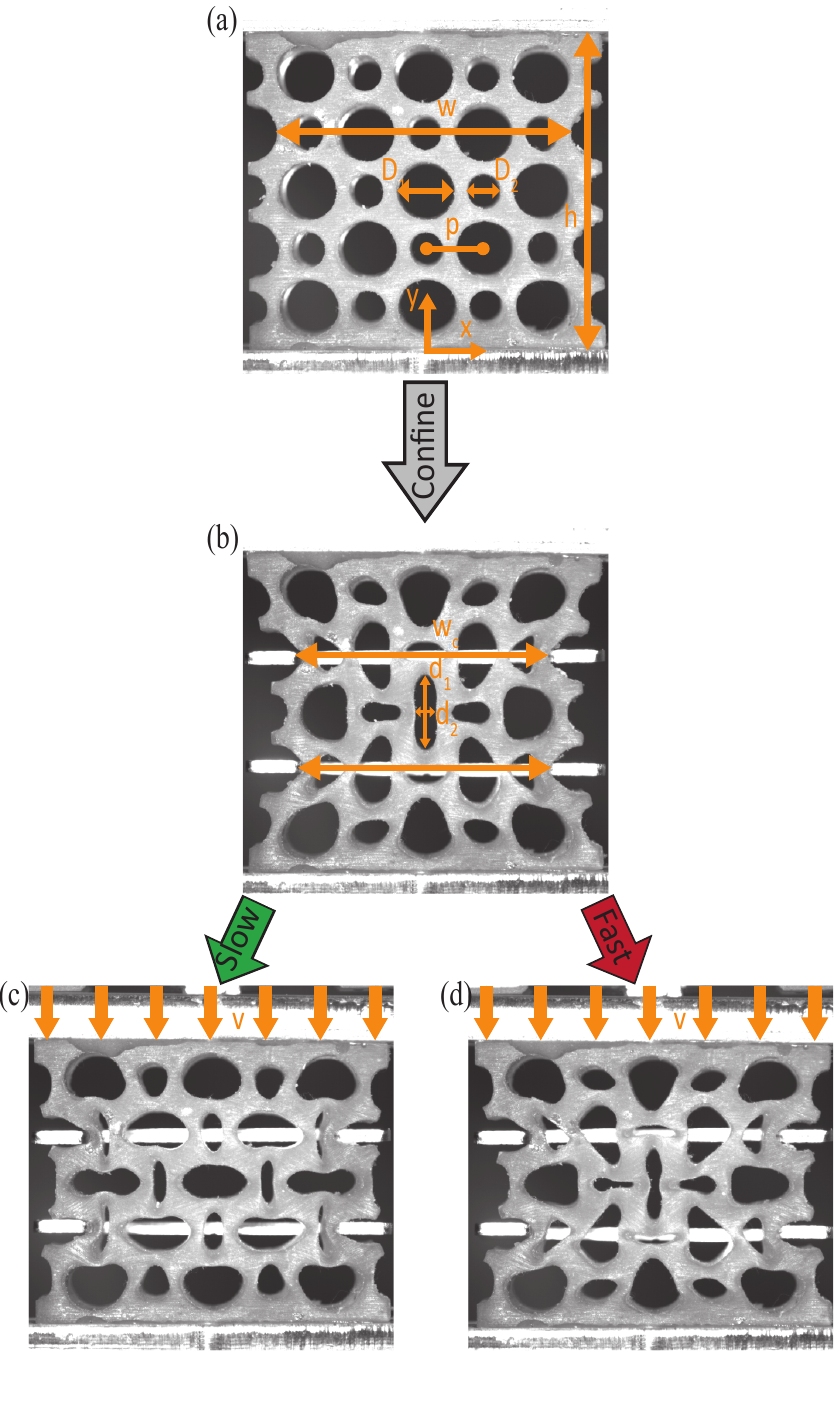}
\caption{(a) Geometry of the metamaterial sample, with $5 \times 5$ alternating holes, characterized by diameters $D_1$ and $D_2$, hole spacing $p$, sample height $h$ and local width $w$. (b) Preconfined with $\epsilon_x = -18 \% $, with the central hole initially in x-polarized position, with central hole major ($d_1$) and minor ($d_2$) axes. (c) Compressing slowly at a rate of $\dot{\epsilon}_y = 9.25 \cdot 10^{-5} \, s^{-1}$  leads to a snap-through instability , changing from x- to y-polarized. Repeating this process fast, with $\dot{\epsilon}_y = 0.3\, s^{-1}$ suppresses the instability, with the sample instead remaining in an x-polarized position.}
\label{fig:Figure1}
\end{figure}

Here, we investigate the role of dissipation in nonlinear snap-through metamaterials. Specifically, we probe how the constitutive materials' viscoelasticity influences the response of a metamaterial, that had been demonstrated earlier to exhibit a programmable hysteric response at slow loading rate when produced from a nearly-ideal elastic rubber \cite{florijn1,florijn2}. We focus on metamaterials consisting of a $5\times 5$ alternating square pattern of circular holes, as seen in Fig. \ref{fig:Figure1}(a), which have been analyzed previously in the quasi-static regime using a near ideally elastic material at low strain rates \cite{florijn1,florijn2}. It was found that when such samples were confined laterally, as shown in Fig. \ref{fig:Figure1}(b), they would exhibit a snap-through response under compression, during which the central hole changes from an x-polarized state in Fig. \ref{fig:Figure1}(b) to a y-polarized state in Fig. \ref{fig:Figure1}(c), inducing a large reduction in force. During unloading, a delayed snap-back instability would occur for lower strain, inducing a large geometrically induced hysteresis.

However, if the sample is made instead of a viscoelastic rubber with a large stress-relaxation effect and we compress the sample quickly, no pattern change is provoked in Fig. \ref{fig:Figure1}(d). This suggests a nontrivial relation between the nonlinear response---induced by the elastic instability---and dissipation---induced by viscoelasticity.

The article is structured as follows. We first describe the sample and production process. Second, we show how we calibrate the 3D printed material using a stress-relaxation test. We then present the experimental results, where we compress the sample using a variety of confinements across a wide range of strain rates, for which we systematically quantify the mechanical hysteresis and geometrical pattern change. In particular we observe an intricate balance between the geometrically induced hysteresis and the viscoelasticity induced hysteresis, such that
optimal dissipation depends both on confinement and applied strain rate in a nontrivial manner. Finally, we present the soft viscoelastic mechanism model and its results, which we use to capture the mechanics of the interaction between viscoelasticity and mechanical instabilities.

Our study opens new a venues for the rational design of viscoelastic metamaterials, whose response drastically changes with loading rate or provide optimal energy absorption performance, combining geometrically induced hysteresis and viscous dissipation.

\section{Sample fabrication, experimental methods and calibration}

We produce the sample in Fig. \ref{fig:Figure1}(a) using additive manufacturing from a rubber-like PolyJet Photopolymer (Stratasys Agilus 30) using a Stratasys Objet500 Connex3 printer. The sample considered has hole sizes $D_1$ = 10mm, $D_2$ = 6 mm, hole pitch $p$ = 10 mm, local width $w$ = 50 mm, height $h$ = 54 mm and thickness $t$ = 35 mm. We define the biholarity, $\chi = \left| D_1 - D_2 \right|/p = 0.3$ \cite{florijn1}. 
The flat top and bottom sides of the sample are then glued to two acrylic plates using 2k epoxy glue that allow for clamping in our uniaxial tensile tester, as performed previously \cite{florijn2}. To prepare the sample for testing: (i) we confine the sample laterally using 2 mm thick CNC machined steel U-shaped clamps as shown in Fig. \ref{fig:Figure1}(b), where we define the strain of confinement as $\epsilon_x = (w_c-w)/w$, where $w_c$ is the distance between the clamps and $w$ is the original local width. We choose only to confine the even rows of our sample ($2^{nd}$ and $4^{th}$), as the odd rows differ in initial width; (ii) we leave the sample unconfined.
\begin{figure}[t!]
\includegraphics[width=1.0\columnwidth]{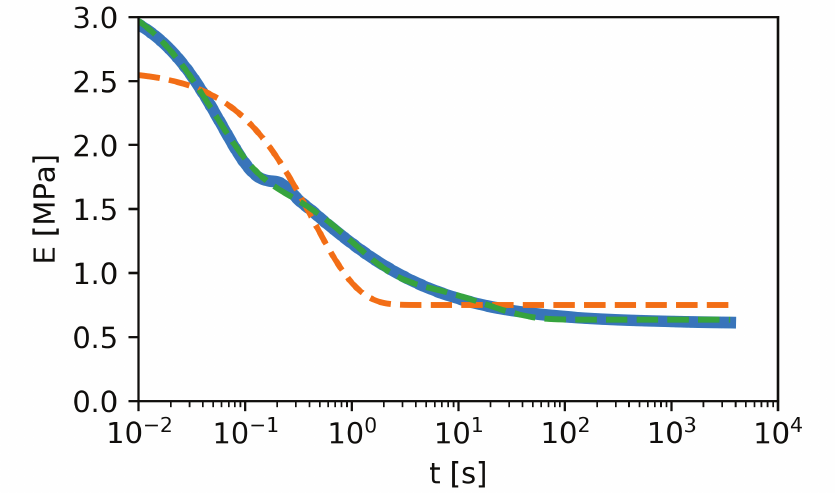}
\caption{Stress-relaxation test for a dogbone sample of Agilus 30 3D printed material. Instantaneous Young's modulus vs. relaxation time. The thick blue line indicates the test results, the thin dashed orange and green curves present fitted data with $N=1$ and $N=3$ respectively.}
\label{fig:Figure_relaxation}
\end{figure}

The sample is compressed uniaxially, both confined and unconfined, using a uniaxial testing device (Instron 5943), which controls the vertical motion of the top side of the sample with a constant controlled velocity v as shown in Fig. \ref{fig:Figure1}(c) and Fig. \ref{fig:Figure1}(d) from $s$ = 0 to $s$ = -12 mm, and immediately unloaded from  $s$ = -12 to $s$ = 0 mm at the same velocity, where we define the loading strain as $\epsilon_y = s/h$ and strain rate $\dot{\epsilon}_y = \left| d\epsilon_y/dt \right|= \left| v/h\right| $. The velocity is controlled down to $\pm$0.1\% of the set speed, while the force is monitored with a 500 N load cell with an accuracy of 0.5\% down to 0.5 N. The load is calibrated to F = 0 N after attaching the sample to the load cell at the top, letting the sample relax for a minute, after which the bottom is clamped to the test bench. The test is further recorded using a high resolution (2048 $\times$ 2048) grayscale CMOS camera (Basler acA2040 with Computar 75 mm lens), inducing a spatial resolution of 0.05 mm. The camera is synchronized with the uniaxial testing device, with the frame rate set such as to produce at least 400 frames per test, with exception of the highest strain rates, where the camera frame rate limits us to extract around 75 frames per test. We extract the position and ellipticity of the central hole by fitting ellipses to the images obtained using standard tessellation techniques. We calculate the polarization, $\Omega$ of the central hole, defined as Eqn. (\ref{eq:polarization}) \cite{florijn1,florijn2}:
\begin{equation}
    \Omega = \pm (1-d_2/d_1)\cos2\phi, 
    \label{eq:polarization}
\end{equation}
where $d_1$ and $d_2$ are the major and minor axes of the ellipse and $\phi$ is the angle between the major axis and the x-axis. The sign of $\Omega$ is fixed such that it is negative in an x-polarized position as in Fig. \ref{fig:Figure1}(b).

 The production process results in a very viscoelastic polymer, for which we characterize the viscoelastic properties using a dogbone stress-relaxation test. In a single process, we 3D printed a sample with a central slender section of Agilus 30 and comparatively rigid top and bottom sides of Stratasys VeroBlackPlus, which are used for clamping in the uniaxial testing device. The rigid sides are used to avoid prestress due to clamping. The Agilus 30 section has a length $L$ = 50 mm, depth $d$ = 5 mm and width $w$ = 10 mm. The sample is stretched quickly to a strain $\epsilon = 20\%$, similar to the global strains induced on the main sample during confinement and compression, at a strain rate of $\dot{\epsilon}=0.4 \, s^{-1}$, after which the force is allowed to relax for one hour. The data is measured with a frequency of 1000 Hz with $t=0 \,s $ defined at the point of highest load.
 
 To describe the stress-relaxation response, we assume the instantaneous time dependent Young's modulus, $E(t)$, can be modelled using a neo-Hookean material model, which can then be characterized using Eqn. (\ref{eq:relax_E}):
 \begin{equation}
 E(t) := \frac{3}{\lambda-\frac{1}{\lambda}} \frac{F}{w d},
 \label{eq:relax_E}
 \end{equation}
 where $\lambda$ is the applied stretch ratio and $F$ is the measured force.
\begin{table}[t!]
\caption{Fitted viscoelastic material properties for Agilus 30, with $N$ = 1 and $N$ = 3}
\begin{center}
\label{tab:relax}
\begin{tabular}{c c c c c c c c }
\hline
$N$ & $E_0$ & $\eta_1$ & $\eta_2$ & $\eta_3$ & $t_1$ & $t_2$ & $t_3$ \\
& [MPa] & & & & [s] &[s] &[s]\\
\hline
1 & 2.58 & 0.71 & & & 0.43 & & \\
3 & 3.25 & 0.45 & 0.26 & 0.10 & 0.047 & 0.97 & 18 \\
\hline
\end{tabular}
\end{center}
\end{table}

 We characterize the viscoelastic properties by assuming a linear Maxwell-Wiechert viscoelastic material model, containing a linear spring in parallel with a number of spring-dashpot Maxwell elements \cite{Wiechert}. The time response of the Young's modulus to a stress-relaxation test is described using Eqn. (\ref{eq:E_relaxation}):
 \begin{equation}
     E(t) := E_0 \left(1- \sum_{n=1}^{N} \eta_n \left(1-e^{\frac{-t}{\tau_n}}\right)\right),
     \label{eq:E_relaxation}
 \end{equation}
 with $E_0$ the peak Young's modulus under instantaneous load, $\eta_n$ the dimensionless relaxation strength, $\tau_n$ the timescale of the individual Maxwell elements \cite{Christensenvisco} and $N$ the number of Maxwell elements considered. For $N$=1 the model reduces to the Standard Linear Solid model (SLS) \cite{lakesvisco}.
 
 \begin{figure}[h!]
\centerline{\psfig{figure=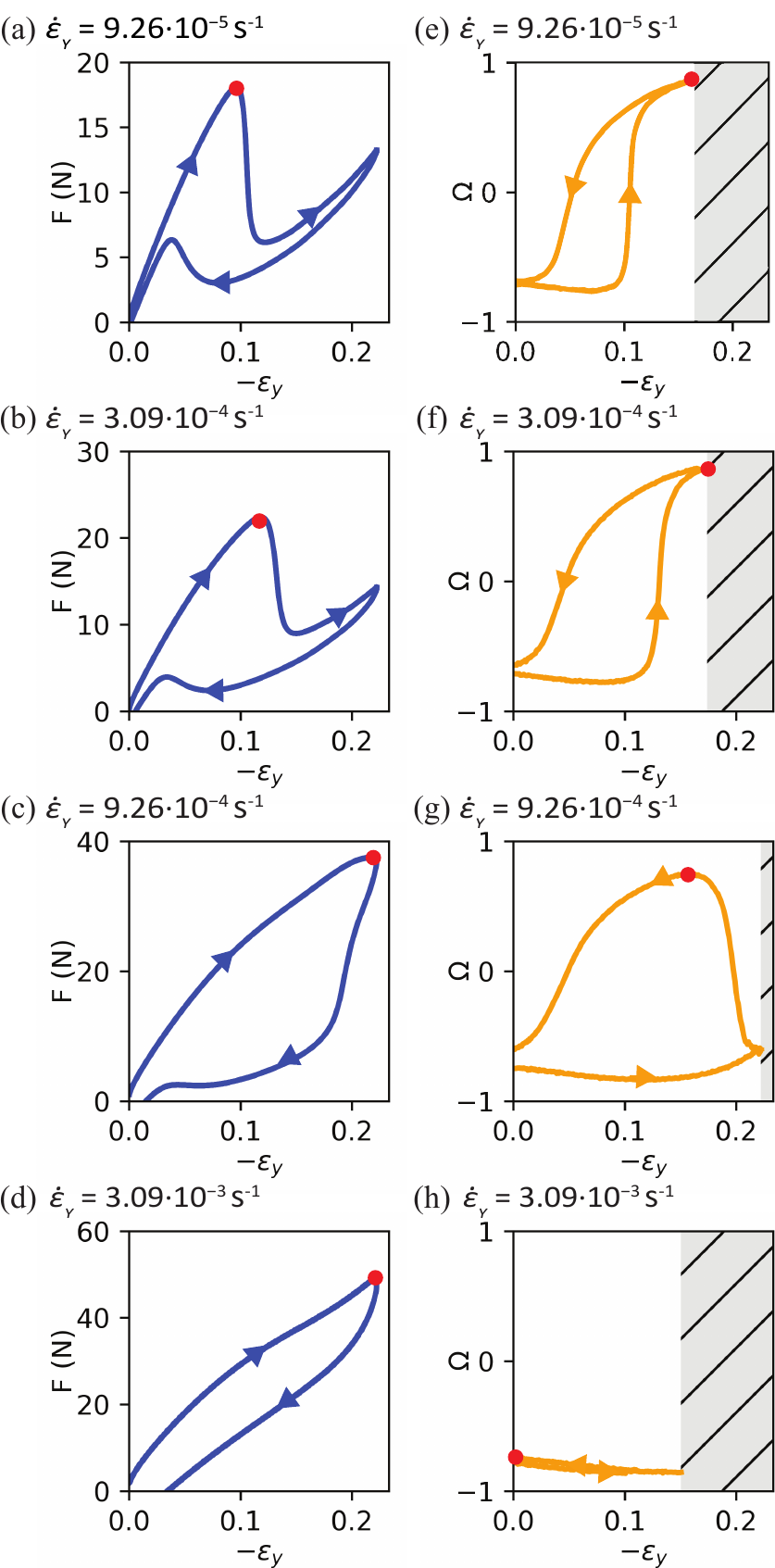,width=.99\columnwidth}}
\caption{Strain rate dependence of the mechanical and geometrical response of the metamaterial. (a)-(d) Force-strain response at increasing compression rates of $\dot{\epsilon}_y = [9.26 \cdot 10^{-5}, 3.09 \cdot 10^{-4},9.26 \cdot 10^{-4},3.09 \cdot 10^{-3}]$ respectively. Red dots highlight the point of maximum force: $(-\epsilon_{y. F_{max}},F_{max})$. (e)-(h) Polarization (see text for definition) vs. strain. Red dots highlight the point of maximum polarization: $(-\epsilon_{y. \Omega_{max}},\Omega_{max})$. Rastered gray areas indicate extreme polarizations which could no longer be detected, because the pore was nearly closed. }
\label{fig:Figure2}
\end{figure}

 The parameters of Eqn. (\ref{eq:E_relaxation}) are then fitted using a least-squares fit to the test data, which was interpolated on a logarithmic time scale from 0.01 s to 1 hour. The relaxation test results are presented in Fig. \ref{fig:Figure_relaxation}, including fits with $N$=1 and $N$ = 3 with corresponding material properties in Tab. \ref{tab:relax}.  Very large viscoelastic effects can clearly be identified. Over the course of an hour, the effective Young's modulus drops by 80\%. A single timescale is clearly insufficient to represent the material response accurately. On the other hand, the material response is modelled accurately using $N$ = 3.

\section{Experimental results}

Figure \ref{fig:Figure1} shows how loading rate can change the response of the metamaterial entirely. The metamaterial in Fig. \ref{fig:Figure1}(a) is first confined laterally  (Fig. \ref{fig:Figure1}(b)). When the sample in Fig. \ref{fig:Figure1}(b) is  compressed slowly, with a strain rate of $\dot{\epsilon}_y = 9.25 \cdot 10^{-5} \, s^{-1}$, a pattern change appears in the central hole, going from x-polarized to y-polarized in Fig. \ref{fig:Figure1}(c), similar to what was previously observed for elastic sample \cite{florijn1,florijn2}. When we repeat this process quickly, with a much higher strain rate of $\dot{\epsilon}_y = 0.3 \, s^{-1}$ in Fig. \ref{fig:Figure1}(d), the material response is entirely different. When compressed, viscous material dissipation induces a time delay between compression and the pattern change such that the sample remains x-polarized over the course of the compression cycle. No pattern change is observed.

In Fig. \ref{fig:Figure2}, we present the force-strain curves of the sample in (a)-(d), for four increasing compression rates from $\dot{\epsilon}_y = 9.26 \cdot 10^{-5}$ to $3.09 \cdot 10^{-3} s^{-1}$. We track the corresponding polarization in (e)-(h). When loaded slowly, as shown in Fig. \ref{fig:Figure2}(a) and Fig. \ref{fig:Figure2}(e), the force-strain curves results in large hysteresis induced by a snap-through instability similar to what was observed previously for elastic materials \cite{florijn1,florijn2}. Namely, the force first increases nearly linearly, which then suddenly drops from 18 N to 6 N at a strain $\epsilon_y= 0.1$ and then increases again until the maximum compression point. Upon decompression, the force first decreases, then snaps back up from 3 N to 6 N at a strain $\epsilon_y= 0.05$,  and finally decreases nearly linearly to zero. 

\begin{figure*}[t!]
\includegraphics[width=6.85in]{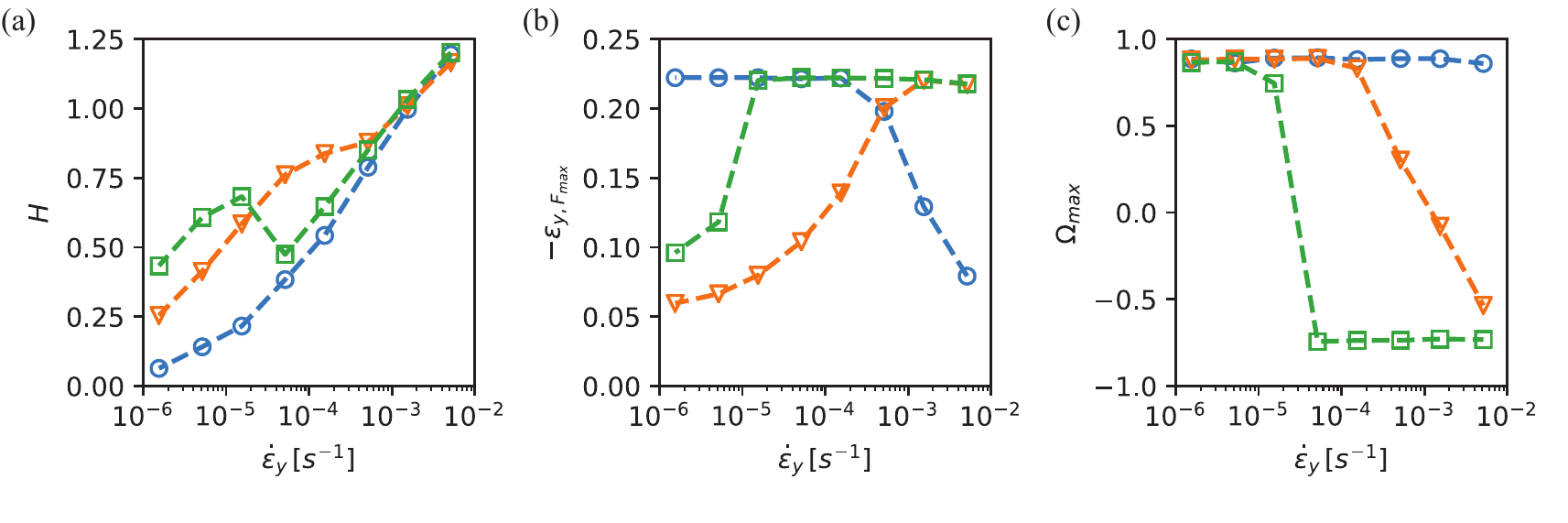}
\caption{Effects of confinement and loading strain rate on the response of viscoelastic metamaterials. (a) Percentage of hysteresis, $H$ vs. strain rate. (b) Strain at which the first local maximum in force is obtained, $\epsilon_{y,F_{max}}$ vs. strain rate. (c) Maximum polarization obtained during loading-unloading, $\Omega_{max}$ vs. strain rate. Legend: blue circles: unconfined, orange inverted triangles: $\epsilon_x=-16\%$, green squares: $\epsilon_x=-18\%$. }
\label{fig:Figure3}
\end{figure*} 

For a larger loading rate ($\dot{\epsilon}_y = 3.09 \cdot 10^{-4}\, s^{-1}$, Fig. \ref{fig:Figure2}(b) and (f)), the effects of viscoelasticity become apparent. First, the overall force scale is larger. Second, the change of polarization is delayed in Fig. \ref{fig:Figure2}(f). Accordingly, the local peak force at instability shifts to a larger strain in Fig. \ref{fig:Figure2}(b), while the peak force of snap-back during unloading shifts to a smaller strain. This leads to an increase of the hysteresis. 

Upon faster loading ($\dot{\epsilon}_y = 9.26 \cdot 10^{-4}\, s^{-1}$), these effects are not only stronger, the response undergoes qualitative changes as well (Figs. \ref{fig:Figure2}(c) and \ref{fig:Figure2}(g)). The force scale is larger, the delay of the pattern change is increased further when loading faster still, to the point where the change in polarization of the central hole is delayed to unloading instead of loading, as seen in Fig. \ref{fig:Figure2}(g). Correspondingly, no local peak force is identified anymore in Fig. \ref{fig:Figure2}(c), while the overall force scale has increased significantly due to strain rate dependent viscoelastic effects. However, the change in polarization during unloading still induces a large amount of hysteresis. 

Finally, for even faster loading rates ($\dot{\epsilon}_y = 3.09 \cdot 10^{-3}\, s^{-1}$) in Fig. \ref{fig:Figure2}(d) and Fig. \ref{fig:Figure2}(h) the instability is entirely suppressed, as shown by the lack of change in the polarization of the central hole in Fig. \ref{fig:Figure2}(h). 
As a result, the hysteresis in Fig. \ref{fig:Figure2}(d) is lower than at slower loading rates, despite the fact that viscoelastic dissipation is larger.

These findings raise the question whether we can identify an optimal combination of geometrically induced and viscoelasticity induced dissipation, depending on how fast the sample is loaded. To obtain more insight on the role of confinement and strain rate, we explore the performance of the metamaterial across a wide range of loading rates for various confinement strains in Fig. \ref{fig:Figure3}. In Fig. \ref{fig:Figure3}(a), we quantify the amount of hysteresis, $H$, defined as the percentage of energy dissipated with respect to the amount of work applied during the loading phase. The strain at which we see a first local maximum force, $\epsilon_{y,F_{max}}$, is presented in Fig. \ref{fig:Figure3}(b). The maximum achieved polarization is given in Fig. \ref{fig:Figure3}(c). Additionally we systematically show data for an unconfined sample that shows no instability as a benchmark (blue circles in Fig. \ref{fig:Figure3}).

For small strain rates $(\dot{\epsilon}_y < 10^{-6})$, geometrically induced hysteresis dominates, where maximum clamping at $\epsilon_x = -18\%$ implies optimal dissipation and almost no dissipation is observed for unconfined samples, as seen in Fig. \ref{fig:Figure3}(a). For the two confined cases considered here, a pattern change is induced as seen in Fig. \ref{fig:Figure3}(c). For confined samples, the accompanying mechanical instability induces a local peak force in Fig. \ref{fig:Figure3}(b). 

As the strain rate increases $(\dot{\epsilon}_y \approx 10^{-5})$, as does the strain rate dependent dissipation. A general increase in hysteresis is seen in Fig. \ref{fig:Figure3}(a), while a delay in pattern change for confined samples shifts the local peak force to a higher strain value as seen in Fig. \ref{fig:Figure3}(b). This was also observed in the analysis of Fig. \ref{fig:Figure2}. 

Interesting results are obtained for higher strain rates $(\dot{\epsilon}_y \approx 10^{-4})$. No more pattern changes occur at the highest level of confinement, as can be seen by the lack of change in polarization for $\epsilon_x=-18\%$ in Fig. \ref{fig:Figure3}(c). This removes the local snap-through peak force for $\epsilon_x=-18\%$ in Fig. \ref{fig:Figure3}(b), which in turn leads leads to a decrease in hysteresis in Fig. \ref{fig:Figure3}(a). As such, optimal dissipation performance is now obtained for a lower level of confinement: $\epsilon_x=-16\%$, for which the pattern change is still obtained. 

As we load faster still $(\dot{\epsilon}_y > 10^{-3})$, strain rate dependent dissipation takes over entirely and hysteresis for all two levels of confinements converge towards the response of the unconfined sample, inducing very large hysteresis exceeding 100\% in Fig. \ref{fig:Figure3}(a), while no more pattern changes are observed for confined samples in Fig. \ref{fig:Figure3}(c). Therefore at large strain rates, the geometrically induced instability is completely suppressed by viscoelastic effects.

\section{Viscoelastic compliant mechanism}

\begin{figure}[t!]
\includegraphics[width=1.0\columnwidth]{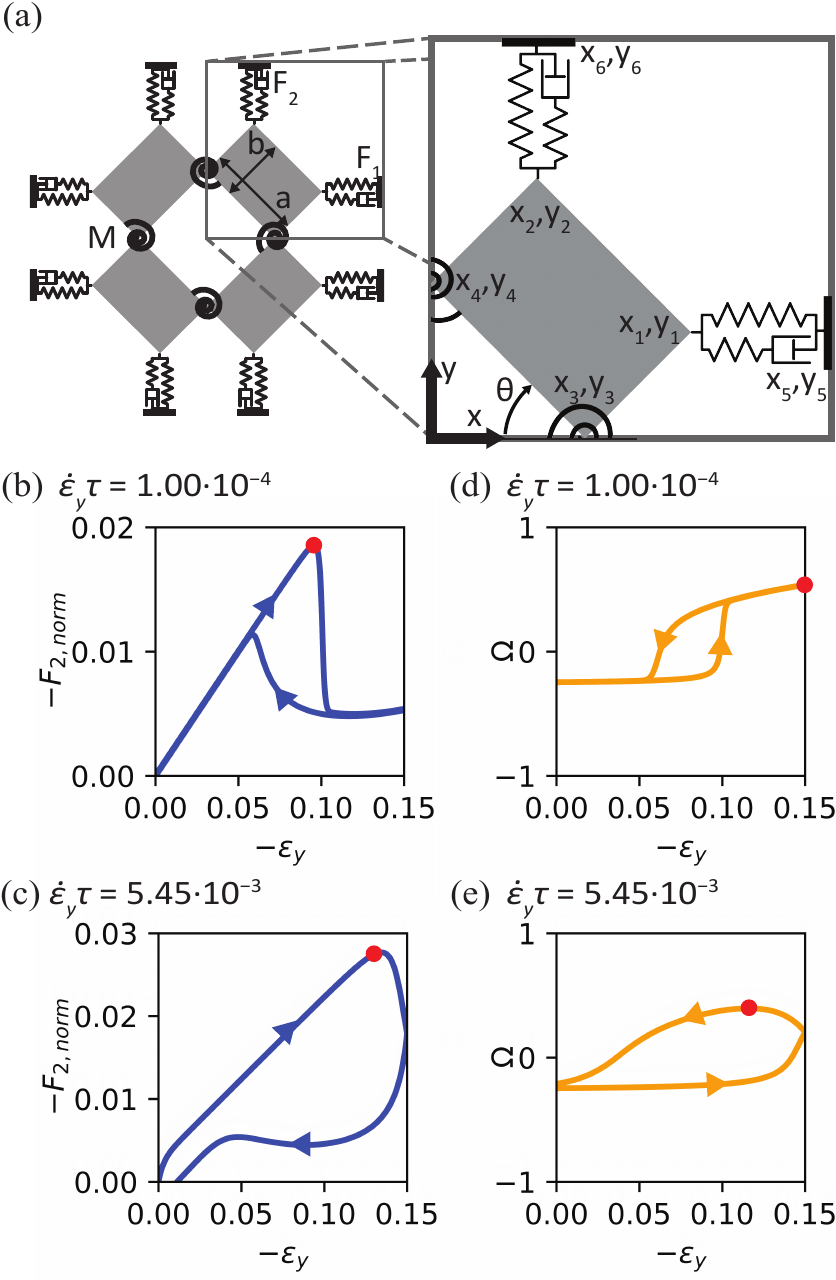}
\caption{Compliant viscoelastic mechanism model. (a) Sketch. Chosen parameters are $\chi = 2(a-b)/(a+b)=0.3$, $\theta_0 = \pi/4$ before confinement, $\eta=0.8$,$c = 4.2\cdot 10^{-4} k a$. (b),(c) Normalized force-strain curves, with $F_{2,norm}=F_2/(kh)$, for a configuration preconfined with $\epsilon_x = - 6.35\%$ ($\theta_1=53^{\circ}$), loaded at strain rates of $\dot{\epsilon}_y\tau =[ 1.0 \cdot10^{-4},  5.45 \cdot10^{-3}]$ respectively. Red dots highlight the point of maximum force: $(-\epsilon_{y. F_{max}},F_{max})$. (d),(e) Corresponding polarization-strain curves. Red dots highlight the point of maximum polarization: $(-\epsilon_{y. \Omega_{max}},\Omega_{max})$. }
\label{fig:Figure4a}
\end{figure}

\begin{figure*}
\includegraphics[width=6.85in]{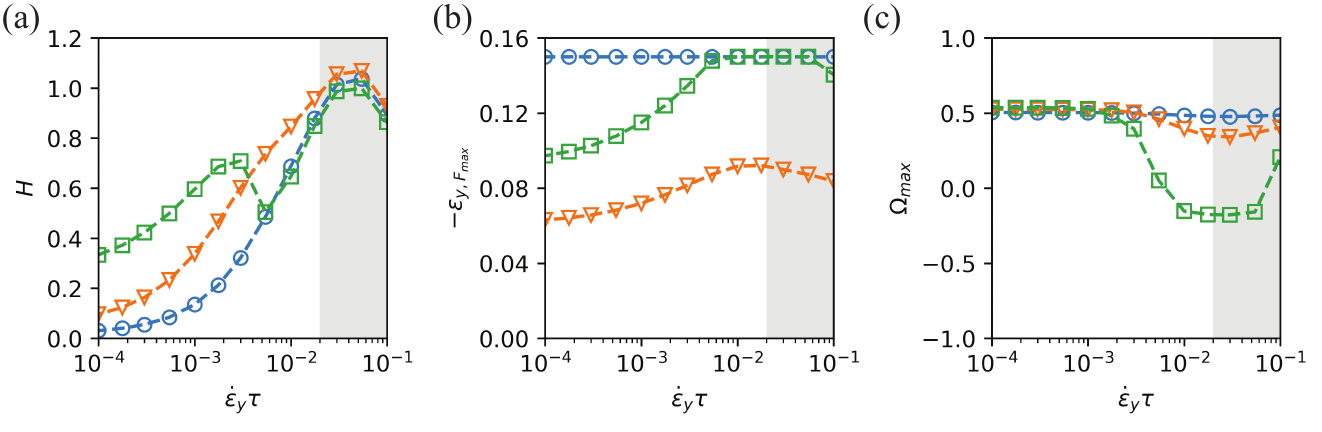}
\caption{Effects of confinement and loading strain rate on the response of the viscoelastic compliant mechanism. (a) Percentage of hysteresis, $H$ vs. strain rate. (b) Strain at which the first local maximum in force is obtained, $\epsilon_{y,F_{max}}$ vs. strain rate. (c) Maximum polarization obtained during loading-unloading, $\Omega_{max}$ vs. strain rate. 
Legend: blue circles: no confinement ($\theta_1=45^{\circ}$), orange inverted triangles: $\epsilon_x=-4.91\%$ ($\theta_1=51^{\circ}$), green squares: $\epsilon_x=-6.35\%$ ($\theta_1=53^{\circ}$). Gray areas indicate strain rates where the model is less valid.}
\label{fig:Figure4b}
\end{figure*}

To get a better qualitative understanding of the mechanics of viscoelastic metamaterials, we enrich the soft mechanism model that was previously introduced to describe the fully elastic response \cite{florijn1}. The confined metamaterials can be represented by a set of rigid rectangles, connected using a set of viscoelastic torsion springs with moment $M$, as seen in Fig. \ref{fig:Figure4a}(a), at an initial angle of $\theta_0=\pi/4$. These rectangles are then connected to the surroundings using viscoelastic springs with an initial length of zero. Symmetry dictates that we only need to model one quarter of the mechanism, i.e. one rectangle, two linear springs and two torsional springs. We pose that the clamps in the x-direction exert a horizontal force $F_1$ and that the time-dependent loading in the y-direction applies a force $F_2$. The model then contains a single degree of freedom, $\theta$, the angle of the rectangle with respect to the horizontal. To account for viscoelastic effects, we restrict our attention to the most basic model for viscoelastic rubbers and use the Standard Linear Solid (SLS) model \cite{lakesvisco}. These ingredients lead to the following constitutive relations:
\begin{eqnarray}
    F_i &=& \dot{u_i}k \tau +u_i    \eta k-\dot{F_i}\tau,
    \label{eq:F_visc}\\
    M &=& 2 \dot{\theta}c \tau +2 \left(\theta-\frac{\pi}{4}\right) c \eta -\dot{M}\tau,
    \label{eq:M_visc}
\end{eqnarray}
Eqn. (\ref{eq:F_visc}) (Eqn. (\ref{eq:M_visc})) apply for the linear (torsional) springs, where all torsional springs induce the same overall moment. 
The subscript $i$ denotes spring 1 or 2, $u$ is the extension of the springs, $k$ and $c$ are the linear and torsional peak stiffness under instantaneous load---comprising both elements of the SLS, in line with the definition of $E_0$ in Eqn. (\ref{eq:E_relaxation})---, $\eta k$ and $\eta c$ are the linear and torsional stiffness after full relaxation---comprising only the elastic part of the SLS---and the relaxation strength, $\eta = 0.8$ is considered, in line with the material stiffness drop of Fig. \ref{fig:Figure_relaxation}. The biholarity, $\chi = 0.3$, defined in Fig. \ref{fig:Figure4b} is used, similar to the experiments. All results are normalized in the time domain with respect to the viscoelastic timescale, $\tau$. Mass is neglected because the experiments are heavily overdamped.
A torsional damper, with negligible torsional damping $\gamma = 7.5\cdot 10^{-10}\cdot k/\dot{\epsilon}_y$ is added to ensure numerical stability. We have checked that changing the value of $\gamma$ by $\pm50 \%$ affects the force response by less than 0.1\%. 

From moment equilibrium, we derive the following constitutive relation:
\begin{equation}
    \dot{\theta} = \frac{F_2\left(x_3-x_2\right) - F_1\left(y_4-y_1\right) -2 M}{\gamma},
    \label{eq:M_eq}
\end{equation}
where coordinates $x_2$, $x_3$, $y_1$ and $y_4$ are defined in Fig. \ref{fig:Figure4a}(a). 

The sides are confined with strain $\epsilon_x$ in the x-direction, after which we can solve the relaxed equilibrium state, inducing an initial confined angle $\theta_1$. The model of Fig. \ref{fig:Figure4a}(a) can only experience a decrease in $\theta$---and therefore a snap-through instability---during loading if $x_2<x_3$, implying we can calculate the maximum confined angle $\theta_{1,max}$:
\begin{equation}
    \theta_{1,max}= \frac{\pi}{2}-\tan^{-1}\frac{b}{a} = 53.5^{\circ}.
    \label{eq:theta1max}
\end{equation} 
Similar to the experiments, we analyze the model for 3 cases: $\theta_1 = 53^{\circ}$ ($\epsilon_x=-6.35\%$, large hysteresis with $\theta_1$ just below $\theta_{1,max}$), $\theta_1 = 51^{\circ}$ ($\epsilon_x=-4.91\%$, intermediate case with less geometric hysteresis) and $\theta_1 = 45^{\circ}$ ($\epsilon_x=0$, unconfined).

The mechanism is subjected to a strain field $\epsilon_y(t)$, bilinear in the time domain, from the top, similar to what was done in the experiments. Namely, it increases linearly until full compression at maximum applied strain of $\epsilon_y=-0.15$ and immediately decreases linearly until zero strain. The response of the system is solved numerically in the time domain using an iterative $4^{th}$ order forward Taylor series method, featuring at least 100 000 time steps, which we have checked is sufficient for numerical convergence.

For a small normalized loading rate, with $\dot{\epsilon}_y\tau = 1.0 \cdot10^{-4} $, the force-strain curve is given in Fig. \ref{fig:Figure4a}(b), with the corresponding polarization in Fig. \ref{fig:Figure4a}(d). Qualitatively, this response is very similar to that of Fig.  \ref{fig:Figure2}(a) and Fig. \ref{fig:Figure2}(e), showing the model matches the experimental results for small quasi-static loading rates. When we increase the loading rate to $\dot{\epsilon}_y\tau = 5.45 \cdot 10^{-3}$, as seen in Fig. \ref{fig:Figure4a}(c) and Fig. \ref{fig:Figure4a}(e) for force-strain and polarization respectively, we see that the displacement at which we achieve maximum force shifts to higher strains, while the change in polarization is delayed, such that it still increases during unloading. The behavior matches qualitatively well the experimental results of Fig. \ref{fig:Figure2}(c) and Fig. \ref{fig:Figure2}(d). Despite its limited complexity, this model can describe the experimental results well and provides illuminating physical insights on the qualitative change of response for larger loading rates as well. During the snap-through event, internal rotations of the mechanism---illustrated by $\Omega$ (\ref{fig:Figure2}(d-e))---are delayed by viscoelastic damping, while the external strain rate remains constant. When such delay becomes longer than the timescale of the experiment, the snap-through instability is suppressed. The only behavior of Fig. \ref{fig:Figure2} which cannot be described is closing of the hole before shape change at high loading rates in Fig. \ref{fig:Figure2}(d) and Fig. \ref{fig:Figure2}(h). We presume that this is due to the fact that the rectangles of Fig. \ref{fig:Figure4a}(a) are considered rigid. Additional degrees of freedom, other than $\theta$ would need to be considered to describe this behavior, which is beyond the scope of the present paper. Also, note that the numerical model is less confined ($\epsilon_x = -6.35\%$) than the experimental sample ($\epsilon_x=-18\%$) to allow for snap-through. We believe that such discrepancy possibly stems from the aforementioned simplifying assumption---wherein deformation of the rectangles is not taken into account---together with finite size effects. 

Having verified that our model qualitatively matches the experiments, we can probe the model further across various loading rates and confinements (Fig. \ref{fig:Figure4b}). Again, the obtained behavior is remarkably similar to the experiments (Fig. \ref{fig:Figure3}). The hysteresis, $H$, in Fig. \ref{fig:Figure4b}(a) shows the same trend as was found in the experiments, where tight confinements induce an increase in energy absorption compared to wider confinements for small loading rates ($\dot{\epsilon}_y\tau < 3 \cdot 10^{-3}$) and the values of the hysteresis, $H$ converge for all confinements at higher strain rates ($\dot{\epsilon}_y\tau \approx  10^{-2}$). Additional confinement can decrease the performance in between these two regimes, as was also observed in the experiments in Fig. \ref{fig:Figure3}(a). 

Figure \ref{fig:Figure4b}(b) also shows results which are similar to Fig. \ref{fig:Figure3}(b) for the experiments for small loading rates ($\dot{\epsilon}_y\tau \approx  10^{-4}$): a local peak force is found for confined samples (indicated by $-\epsilon_{y,F_{max}} < 0.15$), which is induced at a higher strain when confinement increases from 4.91 \% to 6.35 \%. When the loading rate increases ($\dot{\epsilon}_y\tau \approx  10^{-3}$),  $-\epsilon_{y,F_{max}}$ shifts up, which was also observed in the experiments. 

Furthermore, the maximum polarization across confinements and loading rates for the numerical model is given in Fig. \ref{fig:Figure4b}(c). Again, the results are qualitatively very similar to those of the experiments in Fig. \ref{fig:Figure3}(c). The main difference is that the variation of the polarization is smaller in the numerical results. 

However, at high strain rates ($\dot{\epsilon}_y\tau > 2 \cdot 10^{-2}$) all results of the model start to show less correspondence to the experimental results. In fact, the model predicts a decrease in dissipation instead of an increase with increased loading rate and an increase of the maximum polarization for the largest strain rates. This can be attributed to the use of a single viscoelastic timescale in the model: for very large strain rates, the viscous dissipation in the SLS model becomes irrelevant and the response becomes effectively elastic. This shows that results of the model become less valid at higher strain rates.

We can also compare the equivalent timescales of the numerics and experiments. In order to bring the experimental strain rates ($\dot{\epsilon}_y$) of Fig. \ref{fig:Figure3} to the same level as the numerical normalized strain rates ($\dot{\epsilon}_y \tau$) of Fig. \ref{fig:Figure4b}, we need to multiply the experimental strain rates by an equivalent timescale $ \tau_{eq}$, where $1 \cdot 10^{1}\, s < \tau_{eq}< 3 \cdot 10^{2}\, s $.  This is in line with the longest timescale (18 s) obtained during the stress-relaxation test  in Tab. \ref{tab:relax}, providing additional validation for the model.

\begin{figure}
\includegraphics[width=1.0\columnwidth]{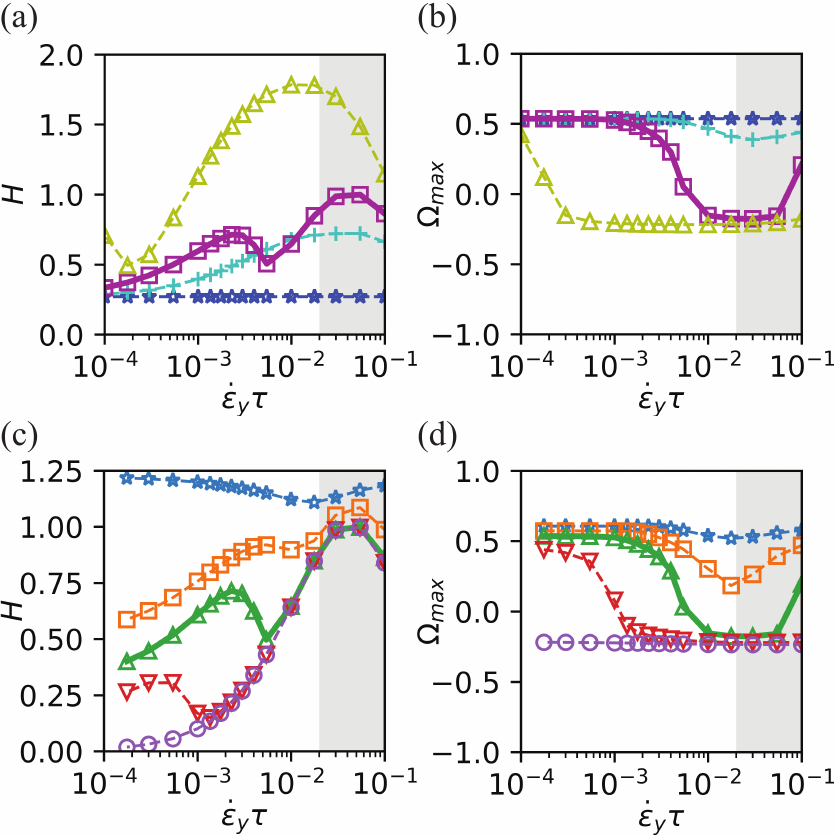}
\caption{Effects of stress relaxation and torsional stiffness and loading strain rate on the viscoelastic compliant mechanism. (a) Percentage of hysteresis $H$ vs. strain rate, for varying viscosities, $\eta$. Legend: blue stars: $\eta=0$, cyan plusses: $\eta=0.5$, magenta squares (thick solid line, reference configuration): $\eta=0.8$, yellow triangles: $\eta=0.99$. (b) Maximum polarization obtained during loading-unloading $\Omega_{max}$ vs. strain rate for the same cases. (c) Percentage of hysteresis $H$ vs. strain rate, for varying torsional stiffnesses, $\beta = c/c_{ref}$, where $c_{ref}$ is the torsional stiffness used for Fig. \ref{fig:Figure4a} and Fig. \ref{fig:Figure4b}. Legend: blue stars: $\beta=0$, orange squares: $\beta = 0.5$, green triangles (thick solid line, reference configuration): $\beta = 1$, red inverted triangles: $\beta = 2$, purple circles: $\beta = 4$.  (d) Maximum polarization obtained during loading-unloading $\Omega_{max}$ vs. strain rate for the same cases. Gray areas indicate strain rates where the model is less valid.}
\label{fig:Fig_model_probe}
\end{figure}

With the viscoelastic model validated, we use the model to probe the design space of viscoelastic metamaterials further. We focus in the following on the influence of relaxation strength and relative torsional stiffness. To do so in a consistent manner, it is important to realize that the preconfining strain necessary to trigger the instability will dramatically change  with the value of the torsional stiffness. We therefore fix the relaxed confined angle $\theta_1 = 53^{\circ}$ instead, the most confined angle analyzed in Fig. \ref{fig:Figure4a} and Fig. \ref{fig:Figure4b}, just below $\theta_{1,max}$. We calculate the equilibrium confinement $\epsilon_x$ accordingly. 
In Fig. \ref{fig:Fig_model_probe}(a) and Fig. \ref{fig:Fig_model_probe}(b), we can see how the relaxation strength influences the hysteresis and pattern change (indicated by $\Omega_{max}$). 

For purely elastic materials ($\eta  = 0$), the pattern change and obtained hysteresis are strain rate independent, as expected. 

For moderate stress relaxation ($\eta = 0.5$), an increase  in hysteresis is found for all strain rates, increasing with loading strain rate until results can no longer be modelled reliably ($\dot{\epsilon}_y\tau > 2 \cdot 10^{-2}$). At $\dot{\epsilon}_y\tau = 2 \cdot 10^{-2}$, a slight decrease in $\Omega_{max}$ is identified.

When the model becomes more viscous ($\eta = 0.8$), namely as in the reference case analyzed above, a shift to lower strain rates is seen, where $H$ and $\Omega_{max}$ are similar to those obtained with $\eta = 0.5$ when the strain rate is an order of magnitude higher. A decrease in viscoelasticity---i.e. with lower dissipation---can be countered with an increase in strain rate---i.e. with higher dissipation. A local minimum in dissipation is observed for $\eta=0.8$  around $\dot{\epsilon}_y\tau = 6 \cdot 10^{-3}$, which was also found in the experiments. This is in contrast with the case $\eta = 0.5$, where no such minimum occurs in the strain rate window considered here.

Energy dissipation increases significantly for materials with extremely large relaxation ($\eta = 0.99$), with $H>1.7$ around $\dot{\epsilon}_y\tau \approx 10^{-2}$. Another shift down in strain rate, with  nearly 2 orders of magnitude, is seen with respect to $\eta = 0.8$. Interestingly, a similar local minimum in dissipation is identified for both $\eta=0.8$ and $\eta=0.99$ ($H \approx 0.5$ for  $\dot{\epsilon}_y\tau \approx 5 \cdot 10^{-3}$ and $\dot{\epsilon}_y\tau \approx  10^{-4}$ respectively), corresponding to the lowest strain rates at which no more pattern change is identified. Furthermore, the geometrically induced hysteresis seems to present a lower limit for hysteresis. Any combination with relaxation strength always induces larger hysteresis, regardless of whether a pattern change occurs.

In a similar way, Fig. \ref{fig:Fig_model_probe}(c) and Fig. \ref{fig:Fig_model_probe}(d) show how torsional stiffness affects the hysteresis and pattern change (indicated by $\Omega_{max}$) in the metamaterial. In general, the hysteresis increases significantly when the torsional stiffness is reduced. When the torsional stiffness is reduced, pattern changes can be obtained at larger strain rates. Furthermore, the elastic energy in the torsional springs during pattern change is reduced relative to the geometric hysteretic energy, inducing a higher percentage of hysteresis overall. These two effects imply an increase in relative dissipation for lower torsional stiffnesses. However, in cases without pattern change ($\beta \geq 1, \dot{\epsilon}_y\tau > 10^{-2}$), the values of the hysteresis, $H$ converge for all confinements. In this regime, dissipation induced by geometrical effects becomes irrelevant and viscoelastic dissipation completely dominates.


\section{Discussion}
In this paper, we have investigated the interaction between elastic snap-through energy dissipation and viscoelastic energy dissipation in mechanical metamaterials. To this end, we 3D-printed, from a photopolymer exhibiting very large stress-relaxation (80\%), confined reprogrammable metamaterials, for which it was previously demonstrated to exhibit an elastic hysteretic response when compressed in a quasi-static manner \cite{florijn1,florijn2}. 

We found that viscoelasticity can change the geometric response of the metamaterial entirely under moderate to large strain rates. A pattern change resulting from a snap-through instability could be delayed or even suppressed when the metamaterial is loaded more quickly, drastically affecting the observed hysteresis. This qualitative change originates from the competition of the timescales of external loading---depending on external strain rate---and internal rotations---mediated by viscoelastic damping. For high strain rates, the energy dissipation is governed entirely by the inherent material dissipation, while for low strain rates, the hysteretic response reduces to the elastic quasi-static response. Strong interactions between geometry and viscoelasticity can be found for intermediate loading rates, where increased lateral confinement reduces the strain rate after which no more pattern change can be observed. As a result, wider lateral confinements, which would provide less dissipation for purely elastic samples, can provide more dissipation for certain moderate strain rates than more narrow confinements would. We also demonstrated that we can capture the behavior of these samples using a simple viscoelastic compliant mechanism model, which we use to explore the design space further. We find that the response of the viscoelastic metamaterial also depends strongly on the elastic hinges connecting the mechanism, which are stressed during the pattern change, as well as the relaxation strength of the constitutive material. Decreasing either the torsional stiffness of the hinges or the relaxation strength or the viscoelastic timescales of the constitutive material can enable dissipation enhancing pattern changes at higher strain rates. Furthermore, we demonstrate that a single viscoelastic timescale is insufficient to describe the response of these metamaterials at higher strain rates.

We therefore show that an optimal energy dissipating metamaterial is not merely a combination of the ideal geometric and viscoelastic dissipative metamaterial. Instead, for any given loading rate, the ideal combination of geometrically induced and viscoelastic dissipation should be selected, still allowing for a mechanical instability. The current metamaterial design we present allows for programming the ideal response for a given loading rate, implying in principle that we can rationally tune the shock damping performance without changing the base material.

Our findings bring a novel understanding of metamaterials in the dynamical regime and opens up avenues for the use of metamaterials for vibration and impact applications, utilising both geometrically induced and viscoelastic dissipation. Furthermore, our findings offer new perspectives for the development of new types of metamaterials which can alter their response in different dynamic situations.

An open question remains how to leverage geometric dissipation at higher strain rates, such that geometrically induced dissipation can still be harnessed, while still harnessing viscous dissipation at lower strain rates, for increased performance across a range of strain rates. The next question is: how general are these results? Is the response under nonlinear vibrations or under shock---at much larger loading rates---similar? Will larger metamaterials and other geometries yield the same results, with or without mechanical snap-through? Finally, the compliant mechanism model only considers a single timescale, which was deemed insufficient at larger strain rates. Will using multiple timescales model result in more reliable results for any loading condition and are three timescales sufficient?

\begin{acknowledgment}
We thank D. Giesen, S. Koot and G. Hardeman for their skilful technical assistance, as well as Jan Heijne at Tata Steel Europe for assistance with mechanical testing. C.C. acknowledges funding from the Netherlands Organization for Scientific Research (NWO) VENI grant No. NWO-680-47-445.
\end{acknowledgment}

%
% ...

\bibliographystyle{Paper}

\bibliography{Paper}

\end{document}